\documentclass[aip,sd,amsmath,amssymb,preprint]{revtex4-1}

\usepackage{graphicx}% Include Fig. files
\usepackage{dcolumn}% Align table columns on decimal point
\usepackage{bm}% bold math
\usepackage{subcaption}

\usepackage[english]{babel}

\begin{document}

%\title{Improving temporal resolution by eliminating RF phase jitter in compression cavities}
\title{Improving temporal resolution of ultrafast electron diffraction by eliminating arrival time jitter induced by radiofrequency bunch compression cavities}

\author{J.G.H. Franssen}
\affiliation{Department of Applied Physics, Eindhoven University of Technology, P.O. Box 513, 5600 MB Eindhoven, The Netherlands}
\affiliation{Institute for Complex Molecular Systems, Eindhoven University of Technology, P.O. Box 513, 5600 MB Eindhoven, The Netherlands}
\author{O.J. Luiten}
\email{o.j.luiten@tue.nl}
\affiliation{Department of Applied Physics, Eindhoven University of Technology, P.O. Box 513, 5600 MB Eindhoven, The Netherlands}
\affiliation{Institute for Complex Molecular Systems, Eindhoven University of Technology, P.O. Box 513, 5600 MB Eindhoven, The Netherlands}

\date{\today}

%%%%%%%%%%%%%%%%%%%%%%%%%%%%%%%%%%%%%%%%%%%%%%
\begin{abstract}

The temporal resolution of sub-relativistic ultrafast electron diffraction (UED) is generally limited by radio frequency (RF) phase and amplitude jitter of the RF lenses that are used to compress the electron pulses. We theoretically show how to circumvent this limitation by using a combination of several RF compression cavities. We show that if powered by the same RF source and with a proper choice of RF field strengths, RF phases and distances between the cavities, the combined arrival time jitter due to RF phase jitter of the cavities is cancelled at the compression point. We also show that the effect of RF amplitude jitter on the temporal resolution is negligible when passing through the cavity at a RF phase optimal for (de)compression. This will allow improvement of the temporal resolution in UED experiments to well below $100~$fs.

\end{abstract}
%%%%%%%%%%%%%%%%%%%%%%%%%%%%%%%%%%%%%%%%%%%%%%


\maketitle

\section{Introduction}
A successful method to improve the temporal resolution in sub-relativistic pump-probe ultrafast electron diffraction (UED) experiments is the use of a resonant radio frequency (RF) cavity in TM$_{010}$ mode\cite{Siwick2002,Morrison2014,Chatelain2014,Mancini2012,Sciaini2011a} to compress electron pulses to the $100~$fs range. In this way single-shot UED has been demonstrated with $100$~fs electron bunches\cite{VanOudheusden2007a,VanOudheusden2010a}. To achieve this the phase of the oscillating electro-magnetic field is synchronized\cite{Brussaard2013} to both the pump and photoemission laser. 

However, RF phase instabilities in the synchronization system lead to variations in the arrival time of the electron bunches, thus limiting the temporal resolution of UED experiments to a few $100~$fs\cite{Siwick2002,Morrison2014,Chatelain2014,Mancini2012,Sciaini2011a}. In addition, RF amplitude instabilities may lead to further degradation of the temporal resolution\cite{Li2009}.

This paper theoretically describes how to eliminate the RF phase jitter using two or three TM$_{010}$ cavities, depending on the velocity chirp of the incoming electron beam. If powered by the same RF source and with a proper choice of RF field strengths, phases and distances between the cavities, the combined phase jitter is cancelled at the compression point. The effect of amplitude instabilities can be minimized by operating the compression cavity at a RF phase for optimal (de)compression. In this way the temporal resolution can be improved substantially. 

This paper is organized as follows: First (Sec.~\ref{theorysect}) we will introduce the concept of using a compression cavity as a longitudinal lens and derive its corresponding focal length. Hereafter (Sec.~\ref{theorysect}A) we will show how RF phase and amplitude fluctuations result in arrival time jitter and how this is connected to the focal length of the longitudinal lens. Next (Sec.~\ref{theorysect}B,C) we show how to use two or three cavities to effectively cancel the arrival time jitter at the compression point. Hereafter (Sec.~\ref{sectGPT}) we will present detailed charged particle tracking simulation results that perfectly agree with the derived analytical theory. We thus show that it is possible to create a longitudinal focus that is inherently insensitive to \emph{both} phase \emph{and} amplitude fluctuations of the RF field in the compression cavities. Finally (Sec.~\ref{seclimit}) we discuss the limitations.

\section{Theory}\label{theorysect}

The principle of using resonant RF cavities as longitudinal lenses for sub-relativistic UED is an established technique which is described in Ref.~\cite{Pasmans2013,VanOudheusden2007a}. The on-axis oscillating electric field inside the RF cavity is given by $\vec{E}=E(z) \cos(\omega t + \phi)~\hat{z}$ with $E(z)$ the on-axis longitudinal electric field amplitude, $\omega$ the angular frequency and $\phi$ the RF phase. The change in longitudinal momentum $\Delta p_{z}$ an electron acquires by traveling trough an RF cavity is given by\cite{Pasmans2013,VanOudheusden2007a}

\begin{equation}
\Delta p_{z} \cong -\frac{eE_{0}d_{c}}{v_{z}} \left(\frac{\omega\zeta}{v_{z}}\sin(\phi)+\cos(\phi)\right),
\label{dpzeq}
\end{equation}

with $e$ the electron charge, $d_{c}=\int_{-\infty}^{\infty} \frac{E(z)}{E_{0}}\cos\left(\frac{\omega z}{v_{z}}\right)dz$ the effective cavity length, $E_{0}=E(0)$ the electric field strength at the center of the cavity, $v_{z}$ the average speed of the electron bunch and $\zeta\equiv z-v_{z}t$ the longitudinal electron coordinate with respect to the center of the bunch; $\phi$ is chosen as the RF phase at the moment the center of the electron bunch passes trough the center of the cavity. The longitudinal focal length $f$ of a such a cavity is given by\cite{Pasmans2013,VanOudheusden2007a}

\begin{equation}
\frac{1}{f}=\frac{-1}{m \gamma^{3} v_{z}} \frac{\partial \Delta p _{z}}{\partial \zeta}=\frac{e d_{c}\omega}{m\gamma^{3}v_{z}^{3}} E_{0}\sin{(\phi)},
\end{equation}

with $m$ the electron mass and $\gamma=1/\sqrt{1-\frac{v^{2}}{c^{2}}}$ the Lorentz factor with $v\approx v_{z}$.

Equation~\ref{dpzeq} shows that the average momentum change $\Delta p_{z}$ of the electron pulse passing through the cavity is zero if the center of the bunch passes through the center of the cavity when the RF electric field goes through zero, i.e. $\phi=\pm \frac{\pi}{2}$. Operating the cavity at a phase of $\phi=\frac{\pi}{2}$ will result in bunch compression: the electrons in the front part of the bunch will be decelerated while the electrons in the back will be accelerated. Operating the cavity at $\phi=-\frac{\pi}{2}$ will result in decompression, the electrons in the front part are accelerated and the ones in the back are decelerated. 

RF phase variations $\delta \phi$ and electric field amplitude fluctuations $\varepsilon \equiv \frac{\Delta E}{E_{0}}$ will result in a net acceleration or deceleration of the electron bunch depending on the sign of $\delta \phi$, $\varepsilon$ and the focal length of the lens. This leads to arrival time fluctuations $\delta t$ at a distance $d$ from the cavity\cite{Pasmans2013}, given by

\begin{equation}
\delta t =\frac{d}{\omega f} \frac{1+\varepsilon}{\tan(\phi+\delta \phi)}
\label{RFjitcond}.
\end{equation}

Equation~(\ref{RFjitcond}) shows that the arrival time depends on the focal length of the lens, so choosing two lenses with opposite focal lengths will allow us to cancel the arrival time fluctuations due to \emph{both} RF phase \emph{and} amplitude fluctuations at some point behind the two cavities. For optimal (de-)compression, i.e. $\phi=\pm \frac{\pi}{2}$, the latter equation reduces to

\begin{equation}
\delta t =-\frac{d}{f} \frac{\delta \phi}{\omega} (1+\varepsilon)\label{dtphaseenergy}
\end{equation}

showing that the arrival time fluctuations due to amplitude fluctuations $\varepsilon$ are a second order effect.

We can illustrate this with a numerical example: state-of-the-art synchronization by an RF phase locked loop system has a typical residual phase RF phase jitter $\delta \phi=2~$mrad\cite{Brussaard2013}. Assuming a typical angular frequency $\omega=2\pi \cdot 3~$GHz and $f=d$ we find $\delta t_{phase}\approx 110~$fs. Solid state RF amplifiers are commercially available with a RF amplitude stability of $\delta P = 5\cdot10^{-4}$, which results in an electric field amplitude stability of $\varepsilon = \frac{\delta P}{2}= 2.5 \cdot 10^{-4}$ and thus to additional arrival time jitter on the order of $\delta t_{amp}=\delta t_{phase} \cdot \varepsilon \approx 28~as$. Clearly the amplitude contribution to the arrival time jitter is negligible, owing to it being a second order effect. The validity of Eq.~\ref{dtphaseenergy} is confirmed by charged particle simulations which will be presented in Section~\ref{sectGPT}. 

RF amplitude fluctuations also causes the longitudinal focus to shift position, thus resulting in bunch length fluctuations at the nominal ($\varepsilon=0$) position of the waist. The Courant-Snyder $\hat{\beta}$ parameter\cite{Courant1958} in the longitudinal waist is given by

\begin{equation}
\hat{\beta}_{waist}=\frac{v_{z}^{2}\tau_{w}^{2}}{\hat{\epsilon}_{z}}\label{Reiglylength}
\end{equation}

with $\tau_{w}$ the pulse length at the longitudinal waist and $\hat{\epsilon}_{z}$ the normalized longitudinal emittance. The Courant-Snyder parameter $\hat{\beta}_{waist}$ is equivalent to the Raighley length in optics. We want $\hat{\beta}_{waist}$ to be much larger than the shift of the focal position to ensure that the pulse length at the nominal focus is not affected by RF amplitude instabilities. This means that the shift in focal position should be much smaller than $\hat{\beta}_{waist}$, i.e. $f \left(1-\frac{1}{1+\varepsilon}\right) < \hat{\beta}_{waist}$, which is equivalent to

\begin{equation}
\varepsilon < \frac{\hat{\beta}_{waist}}{f} = \frac{v_{z}^{2}\tau_{w}^{2}}{f\hat{\epsilon}_{z}}.
\label{Reiglycondition}
\end{equation}

For $100~$keV electrons, $f=0.5~$m, $\tau_{w}=20~$fs and a normalized root-mean-squared (rms) longitudinal emittance $\hat{\epsilon}_{z}=350~$fs$\cdot$eV this results in the condition $\varepsilon < 0.1$, which is easily achievable.

\subsection{Longitudinal focussing}

\begin{figure}[htb!]
\centering
\includegraphics[width=16cm]{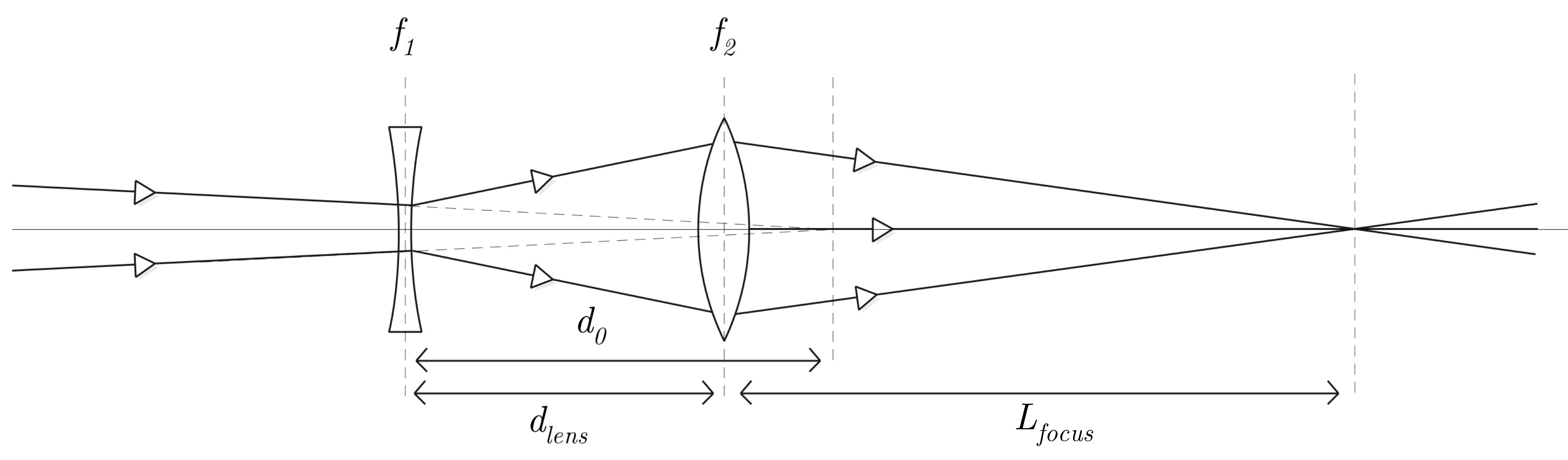}
\caption{Schematic representation of a two-lens longitudinal focussing system. The first lens is a negative lens with a focal length $f_{1}<0$, the second lens is a positive lens with a focal length $f_{2}>0$. The combination of the two lenses compresses the beam at a distance $L_{focus}$ behind the second lens.}
\label{lenssystem}
\end{figure}

We will now first derive how to longitudinally compress an electron pulse by using a two lens focussing system, as is illustrated in Fig.~\ref{lenssystem}. We will use geometrical optics to describe the longitudinal focussing system, i.e. the paraxial beam approximation and thin and weak-lens approximations\cite{Pasmans2013}. 

The first lens is a negative lens with focal length $f_{1}<0$. This lens stretches the electron bunch; the second lens is a positive lens with a focal length $f_{2}>0$. This lens is used to compress the electron bunch, as illustrated in Fig~\ref{lenssystem}. The distance between the lenses is given by $d_{lens}$. The longitudinal divergence of the incoming electron beam is parameterized by the length $d_{0}$, which is the distance of the focal point with respect to the position of the first lens if $\frac{1}{f_{1}}=\frac{1}{f_{2}}=0$. 

$d_{0}>0$ corresponds to a converging beam which is longitudinally focused a distance $d_{0}$ behind the first cavity, as is schematically indicated in Fig.~\ref{lenssystem}. $d_{0}<0$ represents an diverging electron beam which originates from a beam waist a distance $d_{0}$ before the first lens.  

The distance $L_{focus}$ with respect to the second lens (see Fig.~\ref{lenssystem}), of this two-lens system is given by

\begin{equation}
\frac{1}{L_{focus}(d_{0})}=\frac{1 + \varepsilon}{f_{2}}-\frac{1}{d_{lens}-\frac{f_{1}d_{0}}{f_{1}+d_{0}(1+\varepsilon)}}
\label{lfocus}
\end{equation}

with $d_{lens},f_{2} > 0$, $f_{1}<0$ and $\varepsilon$ the electric field amplitude jitter which modulates the focal length of the lens. In the case that the incoming beam is parallel the latter equation reduces to 

\begin{equation}
L_{focus}(d_{0}=\infty)=\lim_{d_{0}\to\infty}L_{focus}(d_{0}) = \frac{f_{2}(d_{lens}-f_{1})}{d_{lens}-f_{2}-f_{1}}.
\end{equation}

In the case that the first lens collimates the incoming converging beam ($f_{1}=-d_{0}$) the position of the focal point becomes

\begin{equation}
L_{focus}(f_{1}=-d_{0})=f_{2}.
\end{equation}

\subsection{Jitter correction}\label{jittercorrsect}

We assume that both cavities have the exact same phase and amplitude variations since they are driven by the same RF amplified signal. From Eq.~\ref{dtphaseenergy} it then follows that for optimal (de)compression the arrival time jitter of the electron pulse at the second cavity (lens 2) due to the first cavity (lens 1) is given by

\begin{equation}
\delta t_{12}=-\frac{d_{lens}}{f_{1}} \frac{\delta\phi}{\omega}(1+\varepsilon),
\end{equation}

Similarly, the arrival time jitter at a distance $L_{jitter}$ behind the second cavity (lens 2) due to the first cavity (lens 1) is given by

\begin{equation}
\delta t_{1L}=-\frac{d_{lens}+L_{jitter}}{f_{1}}\frac{\delta\phi}{\omega}(1+\varepsilon).
\end{equation}

The arrival time jitter at a distance $L_{jitter}$ behind the second cavity (lens 2) due to the second cavity is given by

\begin{equation}
\delta t_{2L}=-\frac{L_{jitter}}{f_{2}}\left(\frac{\delta\phi}{\omega} + \delta t_{12}\right)(1+\varepsilon),
\end{equation}

There will be no arrival time jitter at a distance $L_{jitter}$ behind the second cavity (lens 2) when $\delta t_{1L} + \delta t_{2L} = 0$ which shows that \emph{both} phase $\delta \phi$ \emph{and} amplitude $\varepsilon$ variations cancel in first order. The point where there is no jitter is given by

\begin{equation}
L_{jitter}=\frac{f_{2}d_{lens}}{d_{lens}(1+\varepsilon)-f_{2}-f_{1}}
\label{ljitter}
\end{equation}

with $d_{lens},f_{2}> 0$ , $f_{1}<0$ and $f_{2}<d_{lens}-f_{1}$ since $L_{jitter}>0$. 

To improve the temporal resolution of UED experiments the no-jitter point $L_{jitter}$ has to overlap with the longitudinal focal point $L_{focus}$. These points overlap when the following equation holds:

\begin{equation}
f_{2}=f_{1}\left(\frac{d_{lens}}{d_{0}}-1\right)+d_{lens}(1+\varepsilon)
\end{equation}

with $d_{0}>d_{lens}$ since Eq.(\ref{ljitter}) requires $f_{2}<d_{lens}-f_{1}$. The point where the pulse length is independent of RF phase fluctuations for all electrons inside a bunch is then given by

\begin{equation}
L_{focus}=L_{jitter}=-d_{0}\frac{f_{2}}{f_{1}}=d_{0}\left(1-\frac{d_{lens}(1+\varepsilon)}{f_{1}}\right)-d_{lens}.
\end{equation}

 This means that overlapping the focal point and the no-jitter point is only possible for beams with negative energy chirp ($d_{0}>d_{lens}>0$). This means that in order to cancel the phase jitter in an RF focussing system with two cavities we need an already focussing electron beam. This is the case for an electron beam which is extracted from a longitudinally extended source such as a laser cooled gas\cite{Claessens2005,Franssen2017}. An electron beam extracted from a photo-emission gun can be negatively chirped using magnetic compression schemes. An RF photo-gun operated at the right phase can also produce longitudinally converging bunches.  
 
RF amplitude variations lead to arrival time variations $\delta t_{foc}$ at the position of the jitter correction point given by
 
\begin{equation}
\delta t_{foc} = \varepsilon (1+\varepsilon) \frac{\delta \phi}{\omega} \frac{d_{lens}L_{jitter}}{f_{1}f_{2}}\cong \varepsilon \frac{\delta \phi}{\omega} \frac{d_{lens}L_{jitter}}{f_{1}f_{2}}.\label{dtenergyjitterpoint}
\end{equation}

Here we see that RF amplitude fluctuations in both RF compression cavities result in only small deviations in arrival time at the focus due to the second order nature of the contribution. In addition, amplitude fluctuations lead to the shifts of the focal position given by

\begin{equation}
\delta L_{foc}  = - \varepsilon \frac{L_{focus}^2}{f_{2}}\left(1+\frac{f_{1}}{f_{2}}\right).\label{theorydlfocenergy}
\end{equation}

As shown in Section~\ref{theorysect}, $\left | \delta L_{foc}\right | < \hat{\beta}_{waist}$ which results in the following condition

\begin{equation}
\varepsilon < \frac{f_{2} \hat{\beta}_{waist}}{L_{focus}^{2}\left(1+\frac{f_{1}}{f_{2}}\right)}.\label{2focusreighly}
\end{equation}

which is easily achievable in practice.

\subsection{Three lens jitter correction}

In the previous section we have shown that it is possible to create a jitter free focus using a set of two RF cavities if the incoming beam is longitudinally \emph{converging}. If the incoming beam is longitudinally \emph{diverging} a set of minimally three RF cavities is required to create a jitter free focus. The derivation is similar to the one described in the previous section and yields a jitter free focal point

\begin{equation}
L_{focus}=L_{jitter}=\frac{f_{3}}{f_{1}f_{2}}\left[f_{1}d_{l1}+d_{0}(d_{l1}-f_{1}-f_{2})\right]
\label{threelenssol}
\end{equation}

with the focal length of the third lens

\begin{equation}
f_{3}=\frac{d_{0} d_{l2} (d_{l1}-f_{1})-d_{0}f_{2} (d_{l1}+d_{l2}-f_{1})-f_{1}f_{2}(d_{l1}+d_{l2})+d_{l1} d_{l2} f_{1}}{d_{0} (d_{l1}-f_{1}-f_{2})+d_{l1} f_{1}}
\end{equation}

and with $f_{1}<0,f_{2}>0,d_{l1}>\frac{d_{l2}f_{2}}{d_{l2}-f_{2}}>0,d_{l2}>f_{2}>0$ because $L_{focus}>0$. Here the first and the second lens are separated by a distance $d_{l1}$ and the second and the third lens by a distance $d_{l2}$.

\section{Particle Tracking Simulations}\label{sectGPT}

\begin{figure}[htbp]
\centering
\includegraphics[width=12cm]{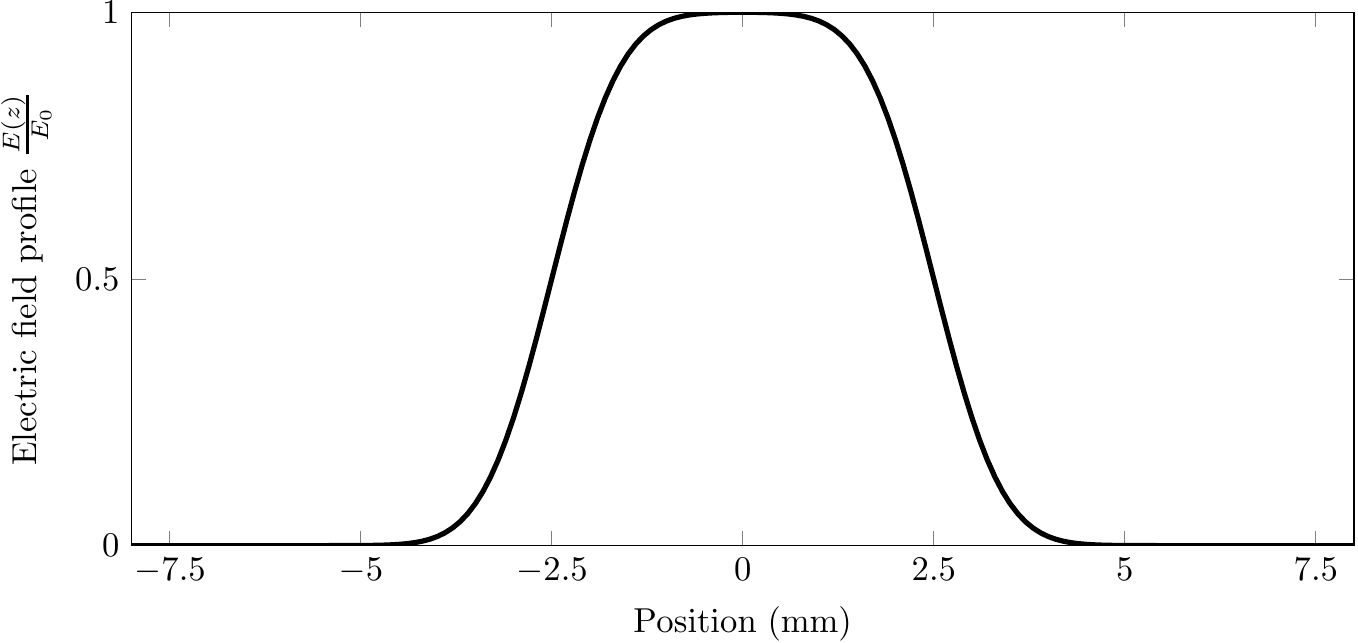}
\caption{The normalized on-axis electric field profile $\frac{E(z)}{E_{0}}$ in the RF cavities.}
\label{Efieldprofile}
\end{figure}

We have performed particle tracking simulations to test our concept in a realistic setting. The General Particle Tracer software package\cite{GPT} was used for calculating the particle trajectories. The full 3D electromagnetic fields inside the RF cavities were calculated by a field expansion\cite{Pasmans2013} using the on-axis normalized field distribution $\frac{E(z)}{E_{0}}$ shown in Fig.~\ref{Efieldprofile}.

In all simulations we used an electron beam with an average beam energy of $100~$keV, a rms transverse emittance $\hat{\epsilon}_{\perp}=30~$pm$\cdot$rad and a normalized rms longitudinal emittance of $\hat{\epsilon}_{z}=2~$ps$\cdot$eV. Space-charge effects have not been taken into account.

\begin{figure}[h]
\centering
\includegraphics[width=12cm]{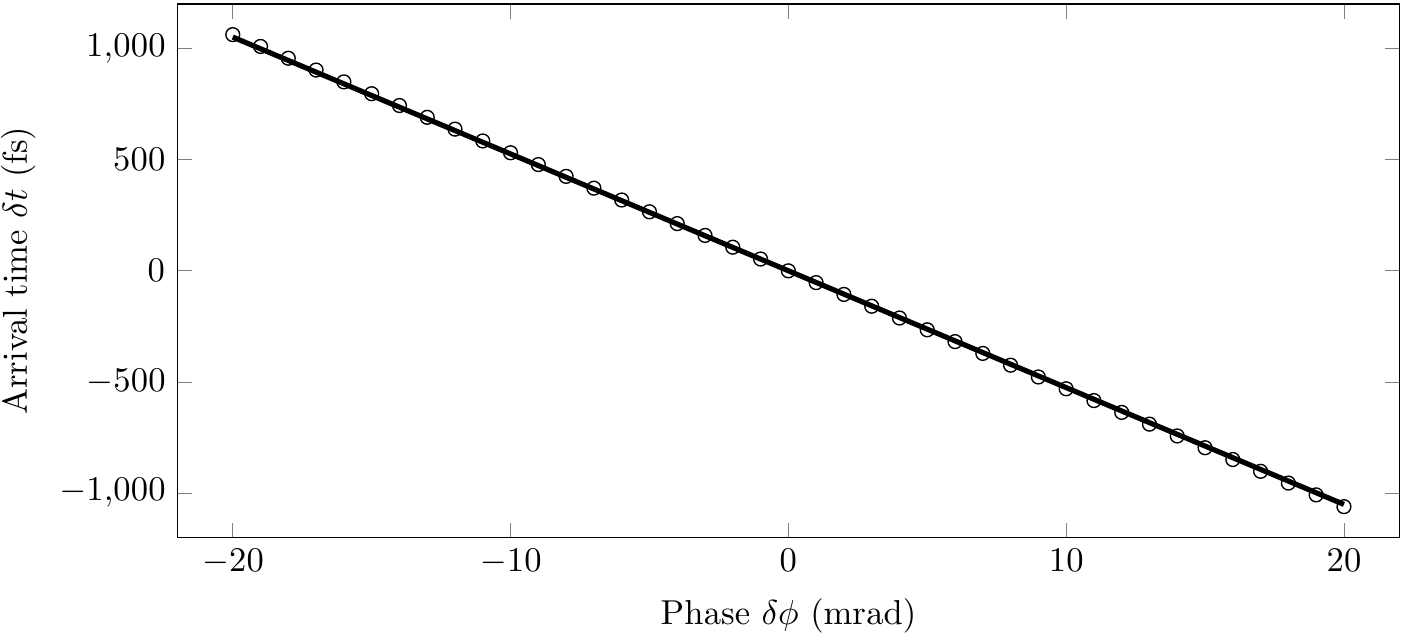}
\caption{Simulation (circles) of the arrival time at the longitudinal focus as a function of the RF phase offset $\delta \phi$. The solid line was calculated using Eq.~\ref{dtphaseenergy} with $\varepsilon=0$.}
\label{Arrivaltimesingle}
\end{figure}

First we simulated the arrival time jitter of a conventional single-cavity focussing system. The electron bunch was longitudinally compressed at distance $L_{focus} \approx f \approx 450~$mm behind the cavity. Figure~\ref{Arrivaltimesingle} shows the arrival time of such an electron bunch at the position of the focus for various RF phase offsets $\delta \phi$. The simulated arrival time is indicated by the circles. The solid line represents the theoretical arrival time jitter (Eq.~\ref{dtphaseenergy}) and perfectly agrees with the simulations. The arrival time jitter follows the linear behavior even beyond $20~$mrad of phase jitter.

\begin{figure}[htbp]
\centering
\includegraphics[width=12cm]{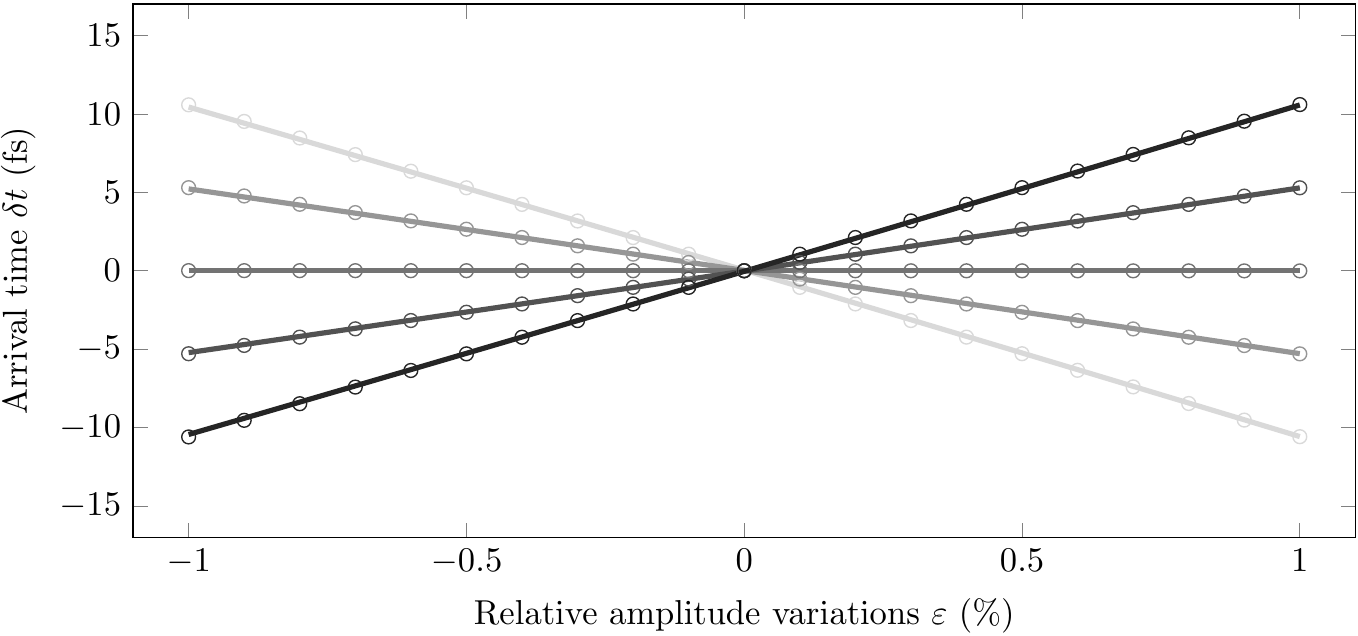}
\caption{Simulation results (circles) of the arrival time with respect to the $\varepsilon=0$ arrival time as a function of $\varepsilon$ for various phase offsets $\delta \phi$ ranging from $\delta \phi=-20~$mrad (black) to $\delta \phi=20~$mrad (grey). The solid lines were calculated using Eq.~\ref{dtphaseenergy}.}
\label{ArrivaltimedE}
\end{figure}

Next we simulated the arrival time dependence on relative electric field amplitude variations $\varepsilon$. Figure~\ref{ArrivaltimedE} shows the arrival time difference for various phase offsets, from $\delta \phi=-20~$ mrad to $\delta \phi=20~$mrad in steps of $10~$mrad. The circles indicate the simulation results and the solid lines are calculated using Eq.~\ref{dtphaseenergy}. Again the theory perfectly describes the simulations, showing that the amplitude fluctuations are indeed a second order effect.

Subsequently we simulated the elimination of the arrival time jitter in the longitudinal focus by using two RF cavities. According to theory (Section~\ref{jittercorrsect}) we can eliminate the RF phase jitter of an already focussing electron bunch (i.e. $d_{0}>0$) with a two lens focussing system, as schematically indicated in Fig.~\ref{lenssystem}. In the simulations $d_{0}=700~$mm, $d_{lens}=200~$mm and $f_{1}=-1000~$mm.

\begin{figure}[htbp]
\centering
\includegraphics[width=12cm]{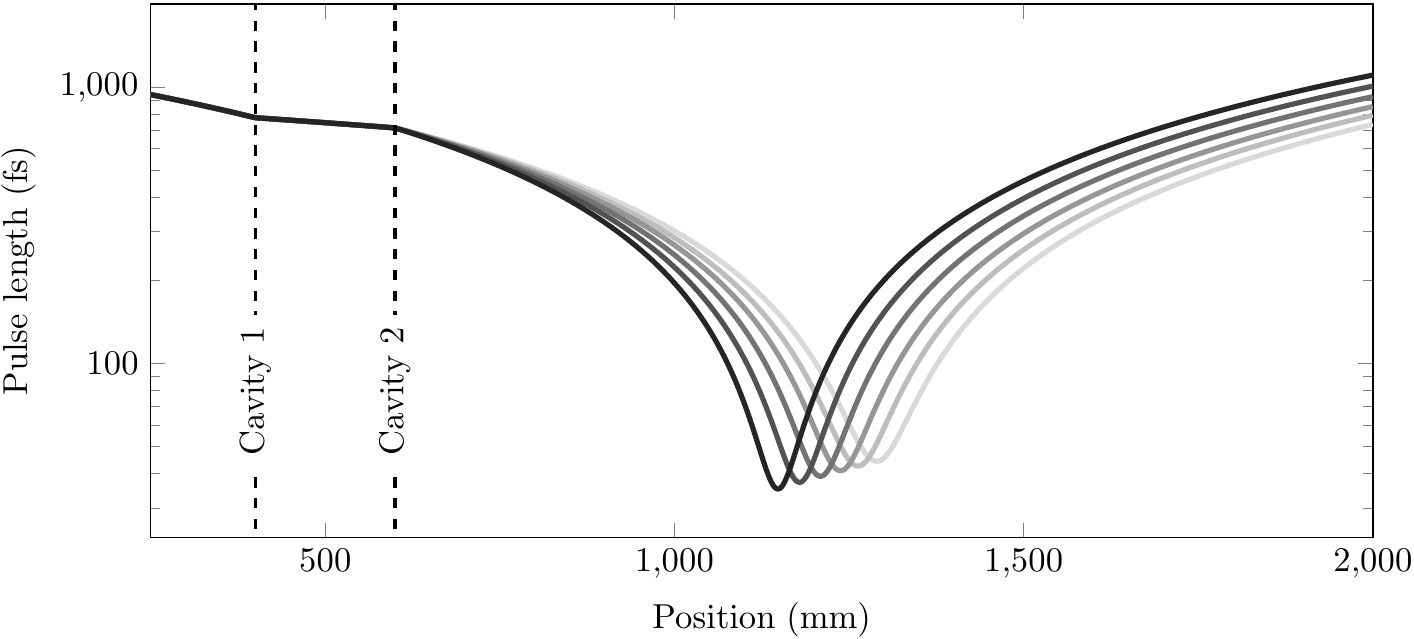}
\caption{Simulated electron pulse length as function of position for focal lengths $f_{2}$ ranging form $f_{2}=750~$mm (black) to $f_{2}=1000~$mm (grey). The dashed lines indicate the positions of the cavities. Cavity 1 decompresses ($f_{1}<0$) the electron pulse and cavity 2 compresses the electron pulse ($f_{2}>0$).}
\label{pulselength2cav}
\end{figure}

Figure~\ref{pulselength2cav} shows the pulse length of the electron bunch as function of longitudinal position for various focal lengths $f_{2}$, ranging from $f_{2}=750~$mm to $f_{2}=1000~$mm in steps of $50~$mm. The dashed lines indicate the positions of the first and second cavity. The figure shows that we enter the first cavity with a negatively chirped bunch. The first cavity defocusses ($f_{1}<0$) the electron bunch; the front gets accelerated and the back decelerated. The second cavity focusses ($f_{2}>0$) the electron beam.

\begin{figure}[htb!]
\centering
\includegraphics[width=12cm]{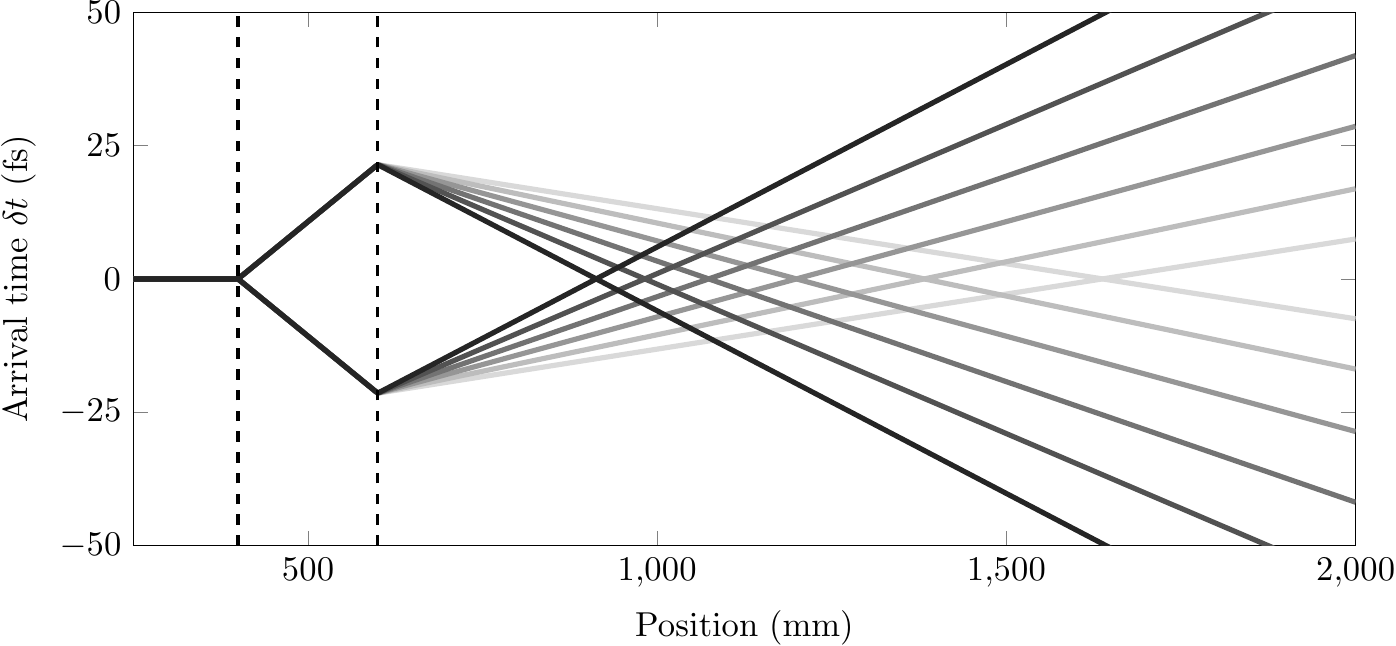}
\caption{Simulated arrival time as a function of position with respect to the $\delta \phi=0$ arrival time for $\delta \phi=\pm2~$mrad and focal lengths $f_{2}$ ranging from $f_{2}=750~$mm (black) to $f_{2}=1000~$mm (grey). The dashed lines indicate the positions of the cavities.}
\label{arrivaltime2cav}
\end{figure}

Figure~\ref{arrivaltime2cav} shows the simulated arrival time with respect to the $\delta \phi=0$ arrival time of an electron bunch which passed trough the cavity with a phase offset $\delta \phi =\pm2~$mrad for focal lengths of the second lens ranging from $f_{2}=750~$mm to $f_{2}=1000~$mm in steps of $50~$mm. At certain positions behind the second cavity the arrival time difference cancels out. The dashed lines again indicate the positions of the first and second cavity.

\begin{figure}[htb!]
\centering
\includegraphics[width=12cm]{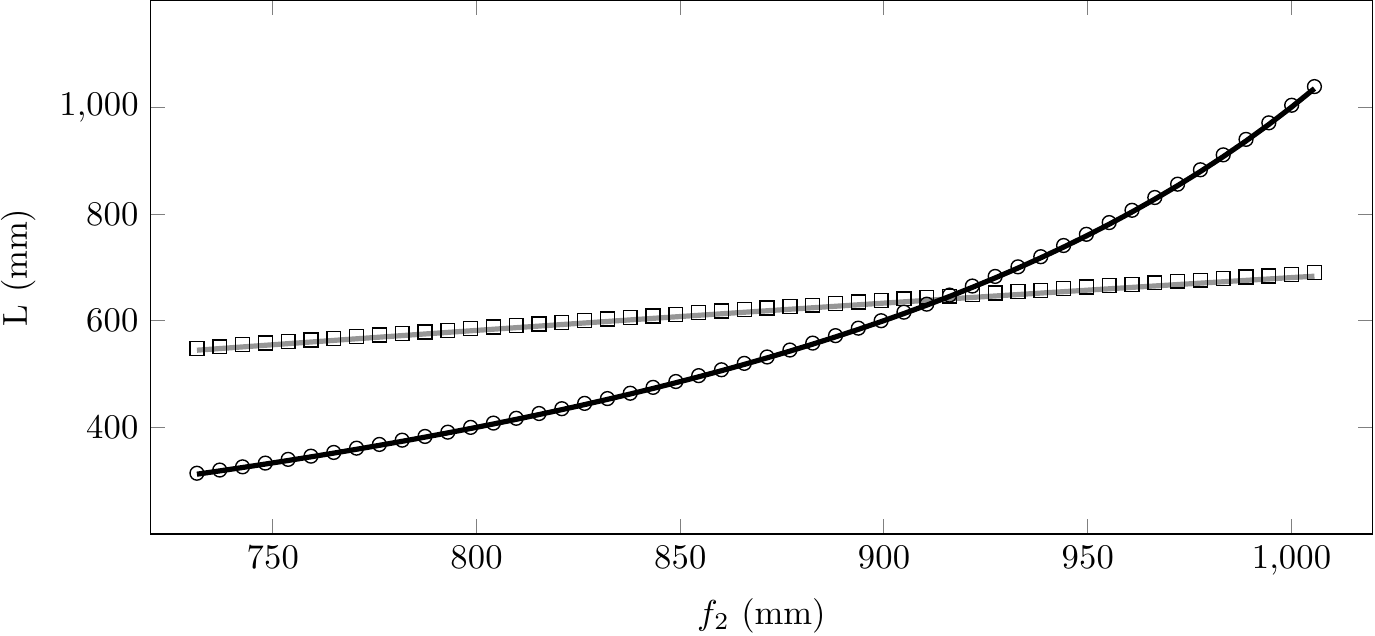}
\caption{Longitudinal focal position (squares) $L_{focus}$ and zero jitter point (circles) $L_{jitter}$ as a function of focal length $f_{2}$. The solid grey curve was calculated using Eq.~\ref{lfocus}; the solid black cure was calculated using Eq.~\ref{ljitter}.}
\label{LfocusLjitter}
\end{figure}

From Fig.~\ref{arrivaltime2cav} we can determine the position of the zero jitter point, $L_{jitter}$. Similarly we can determine the focal point $L_{focus}$ from Fig.~\ref{pulselength2cav}. Figure~\ref{LfocusLjitter} shows both $L_{jitter}$ (circles) and $L_{focus}$ (squares) as a function of the focal length $f_{2}$. The solid black curve was calculated using Eq.~\ref{ljitter} with $\varepsilon=0$. The solid grey curve was calculated using Eq.~\ref{lfocus} with $\varepsilon=0$. The theoretical curves perfectly describe the simulations. At the position where the longitudinal focus and the zero jitter point intersect we find a longitudinal waist that is insensitive to arrival time jitter due to RF phase fluctuations.

\begin{figure}[htb!]
\centering
\includegraphics[width=12cm]{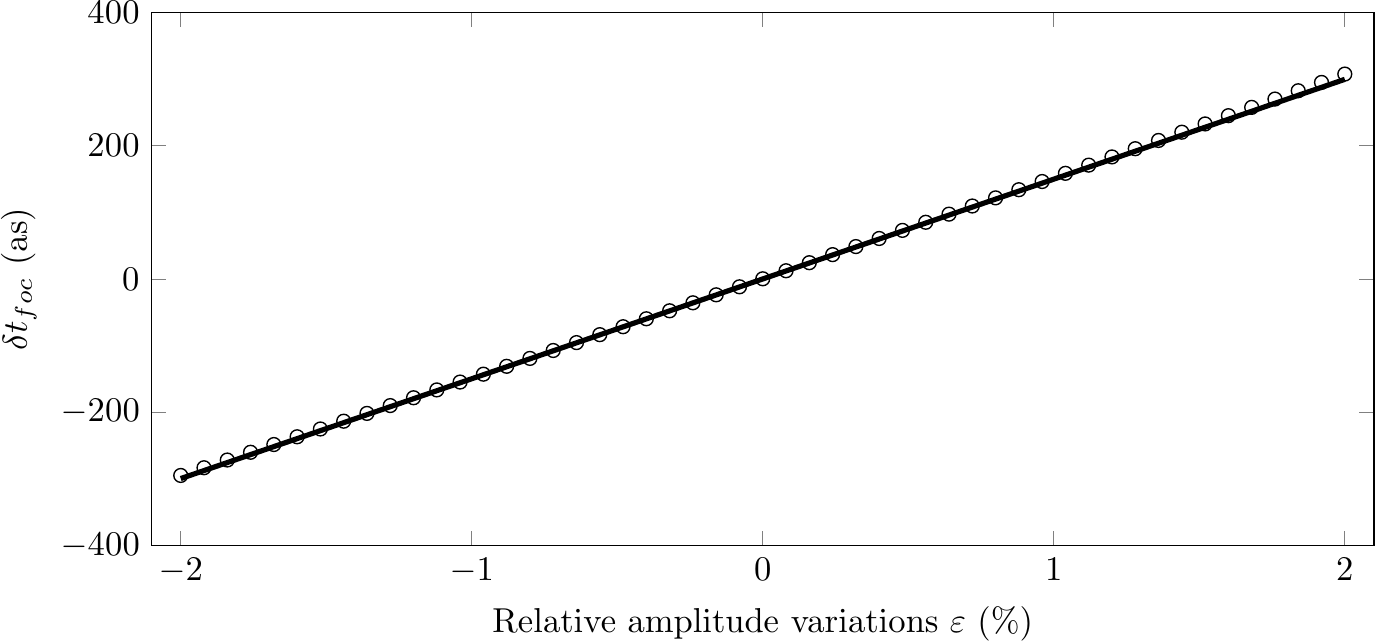}
\caption{Simulated arrival time (circles) at the zero jitter point as a function of the relative electric field amplitude offset $\varepsilon$. The solid line was calculated using Eq.~\ref{dtenergyjitterpoint}.}
\label{dtfocdueenergy}
\end{figure}

Figure~\ref{dtfocdueenergy} shows arrival time at the zero jitter point with respect to the $\varepsilon=0$ arrival time as a function of the relative amplitude variations $\varepsilon$. The circles represent the simulated results and the solid black curve was calculated using Eq.~\ref{dtenergyjitterpoint}. The figure shows that the simulation results are perfectly described by theory, even for electric field amplitude fluctuations up to $|\varepsilon| = 2\%$. The figure also shows that the change in arrival time at the zero jitter point is below half a femtosecond for $|\varepsilon| < 2\%$, which is due to the second order nature of the amplitude fluctuations.

\begin{figure}[htb!]
\centering
\includegraphics[width=12cm]{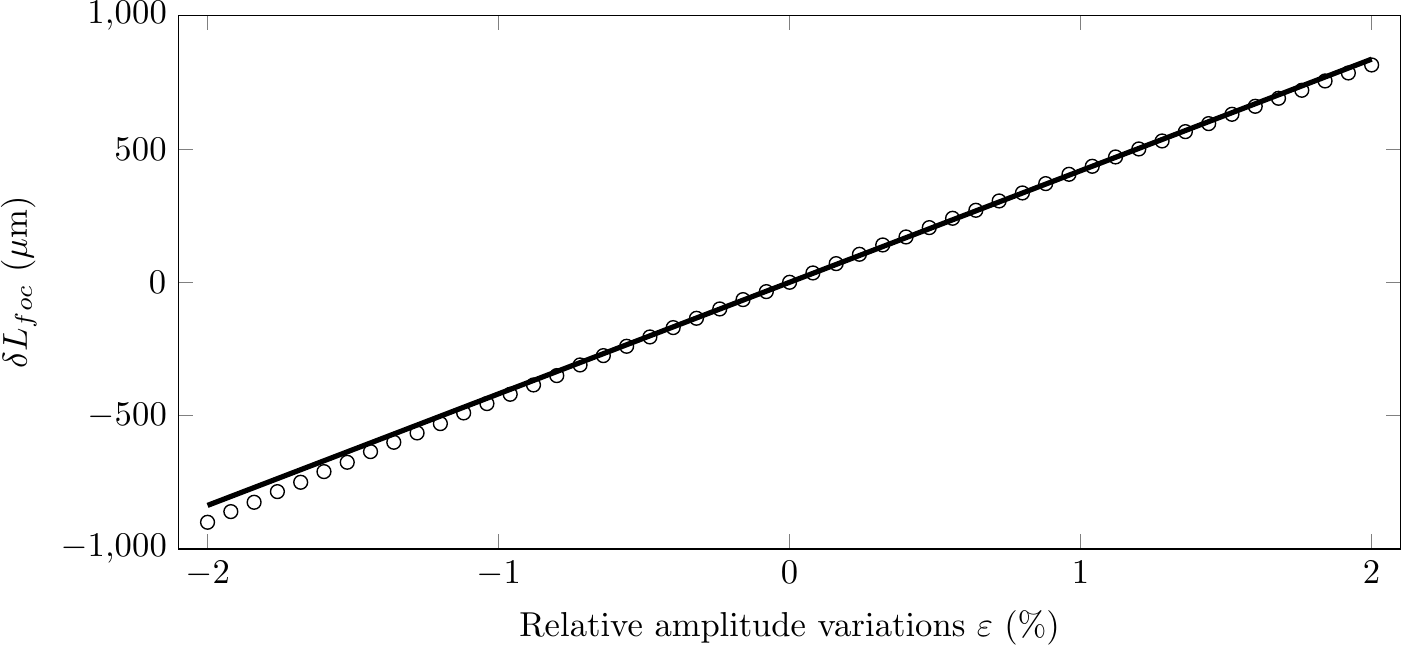}
\caption{Simulated focal position change (circles) as a function of the relative electric field amplitude variations $\varepsilon$. The solid line has been calculated using Eq.~\ref{theorydlfocenergy}.}
\label{dlfocdueenergy}
\end{figure}

Finally, Figure~\ref{dlfocdueenergy} shows the shift of the focal point position as function of relative electric field variations $\varepsilon$. The circles represent the simulation results and the solid black line was calculated by using Eq.~\ref{theorydlfocenergy}. The figure shows that the simulation results are well described by the theory for relative electric field amplitude variations up to $|\varepsilon| = 2\%$. The change of the position of the focal point is below $1~$mm which is much smaller than the typical value of $\hat{\beta}_{waist}$ in a longitudinal focus, which means Eq.~\ref{2focusreighly} is easily satisfied.

We therefore conclude that the analytical theory perfectly agrees with realistic charged particle simulations.

\section{Limitations}\label{seclimit}

The highest frequency $f_{filter}$ of the jitter that can be removed is limited by the time it takes an electron to travel between the cavities: $f_{filter}\leq\frac{v_{z}}{d_{lens}}$. For a $100~$keV bunch this results in $f_{filter}\approx 200$ MHz per meter distance between the cavities.

In this paper we have assumed that the arrival time of the electron bunch at the first RF cavity does not vary in time. Average longitudinal energy fluctuations will result in additional arrival time jitter at the first cavity and thus at the longitudinal focus, limiting the temporal resolution. This will be the limiting factor on the temporal resolution if the arrival time jitter due to RF phase fluctuations is completely cancelled. The arrival time jitter $\delta t_{gun}$ at a distance $d$ from the gun due to relative beam energy fluctuations $\frac{\delta U}{U}$ is given by

\begin{equation}
\delta t_{gun} = \frac{\gamma -1}{\gamma^{3}\beta^{3}} \frac{d}{c} \frac{\delta U}{U}.\label{finaleq}
\end{equation}

As an example, at a distance $d=1~$m, an electron beam energy of $100~$keV and relative energy fluctuations $\frac{\delta U}{U}=10^{-5}$ this results in $\delta t_{gun} = 23~$fs. This is easily achievable for DC photoguns\cite{VanOudheusden2007a}.

For a $1~$MeV electron guns the arrival time fluctuations due to gun jitter will be even lower since $\delta t_{gun}$ scales with $\frac{\gamma-1}{\gamma^{3}\beta^{3}}$. On the other hand the relative energy fluctuations of RF photoguns are larger; in literature\cite{Filippetto2016} a value of $\frac{\delta U}{U}=5 \cdot 10^{-5}$ has been reported, resulting in $\delta t_{gun}= 15~$fs for the same conditions as used above.

This shows that our method should improve the temporal resolution of UED experiments significantly for both sub-relativistic and relativistic UED experiments.

%%%%%%%%%%%%%%%% conclusions %%%%%%%%%%%%%%%%%%%%%%%

\section{Conclusions and outlook}\label{outlookconc}

We have theoretically shown that we can eliminate RF phase jitter in an RF bunch compression system by using a set of two or three RF cavities operated in TM$_{010}$ mode. If powered by the same RF amplifier and with specific values for the distances between the cavities, the focal lengths and the RF phases, the RF jitter can be canceled at the position of the longitudinal focus. If the incoming electron bunch is longitudinally converging, i.e. with a negative chirp, a set of minimally two RF cavities is required. When the incoming bunch is longitudinally diverging, i.e. with a positive chirp, a set of minimally three cavities is required. The analytical theory results are confirmed by charged particle simulations. This means that we can improve the temporal resolution of UED experiments to well below $100~$fs by creating a jitter free longitudinal focus allowing \emph{both} phase \emph{and} amplitude variations.

%%%%%%%%%%%%%%%%%% end %%%%%%%%%%%%%%%%%%%%%%%%%%%%

\begin{acknowledgments}

This research is supported by the Institute for Complex Molecular Systems (ICMS) at Eindhoven University of Technology. 
\end{acknowledgments}

%\bibliography{library.bib}


\end{document}